\documentclass[preprint,12pt]{elsarticle}

\usepackage{amssymb,amsfonts,amsmath}
\usepackage{subfigure}
\usepackage{color} 

\begin{document}

\begin{frontmatter}



\title{Interaction of ultrarelativistic electron and proton bunches with dense plasmas}

\renewcommand{\thefootnote}{\fnsymbol{footnote}}

\author[1]{A. A. Rukhadze}
\author[2]{S. P. Sadykova \footnote{\textcolor{blue}{Corresponding
 author}: Humboldt-Universit\"at zu Berlin, Germany, Newtonstr. 15, 12489 Berlin, Germany;  Tel.: +493020937876, Fax.: +49-30-2093-7638; \textit{E-mail adress}: { saltanat@physik.hu-berlin.de.}}}
\address[1]{Prokhorov General Physics Institute,
Russian Academy of Sciences 119991, Vavilov Str., 38., Moscow, Russia \footnote{\textit{E-mail adress}: rukh@fpl.gpi.ru .} }
\address[2]{Humboldt-Universit\"at zu Berlin, Germany, Newtonstr. 15, 12489 Berlin, Germany}
\begin{abstract}
Here we discuss the possibility of employment of  ultrarelativistic electron and proton bunches for generation of high plasma wakefields in dense plasmas due to the Cherenkov resonance plasma-bunch interaction. We estimate the maximum amplitude of such a wake and minimum system length at which the maximum amplitude  can be generated at the given bunch parameters.  \end{abstract}


\end{frontmatter}


\section{Introduction}
\label{Int}
\renewcommand{\thefootnote}{\arabic{footnote}}\setcounter{footnote}{0}

For the first time in works \cite{1, 2} it was proposed to employ the relativistic electron bunches propogating through plasma for generation of  high plasma wakefields. High-energy bunch electrons generate the wake in such a way that the energy from a bunch of electrons is transferred to the plasma wave  through the Cherenkov resonance radiation producing high electric fields (wakefields). Later, in the work \cite{3} it was shown that the ultrarelativistic electron bunches can generate plasma waves with high relativistic phase velocity including  the transverse electromagnetic waves efficiently radiated out of plasma. This led to the birth of a new applied science - high-power pulse relativistic superhigh-frequency (SHF) electronics \cite{4},\cite{5} enabling to produce the high-power pulse broadened SHF amplifiers and generators on a base of such plasma-bunch systems \cite{6}. \\
	\indent Recently, in works \cite{7,8} the possibility of generation of high power wakefields  (proton-bunch-driven plasma-wakefield acceleration) of terra-watt amplitude using the  ultrarelativistic  proton bunches was introduced.   \\
	\indent  In the present work this idea along with the employment of electron bunch is discussed at the qualitative level. Namely, we make  an estimation of plasma parameters, maximum amplitude of the generated wake when the relativistic electron and proton bunches are used and sysem length at which the maximum amplitude of the wake can be gained.
	
\section{ Ultrarelativistic electron bunch}
\begin{figure}
\subfigure
{\includegraphics[height=6.8 cm,width=7.7cm]{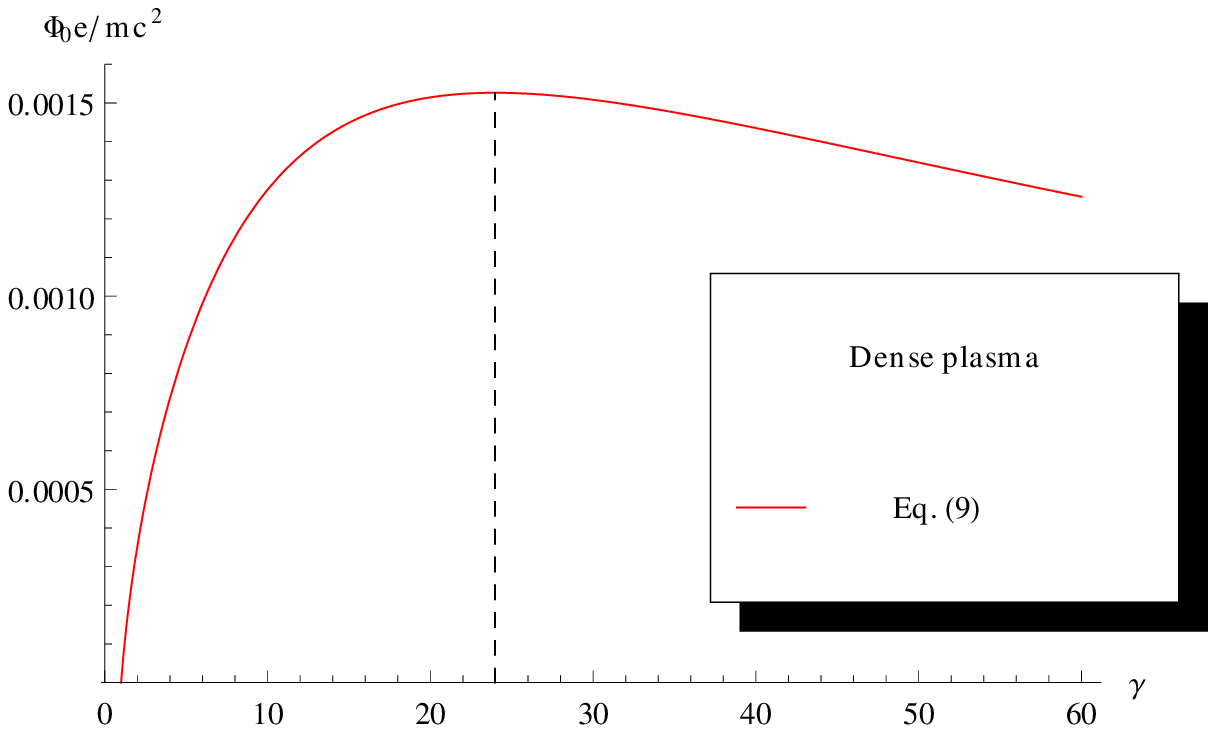}~a)}
\subfigure
{\includegraphics[height=6.8 cm,width=7.6cm]{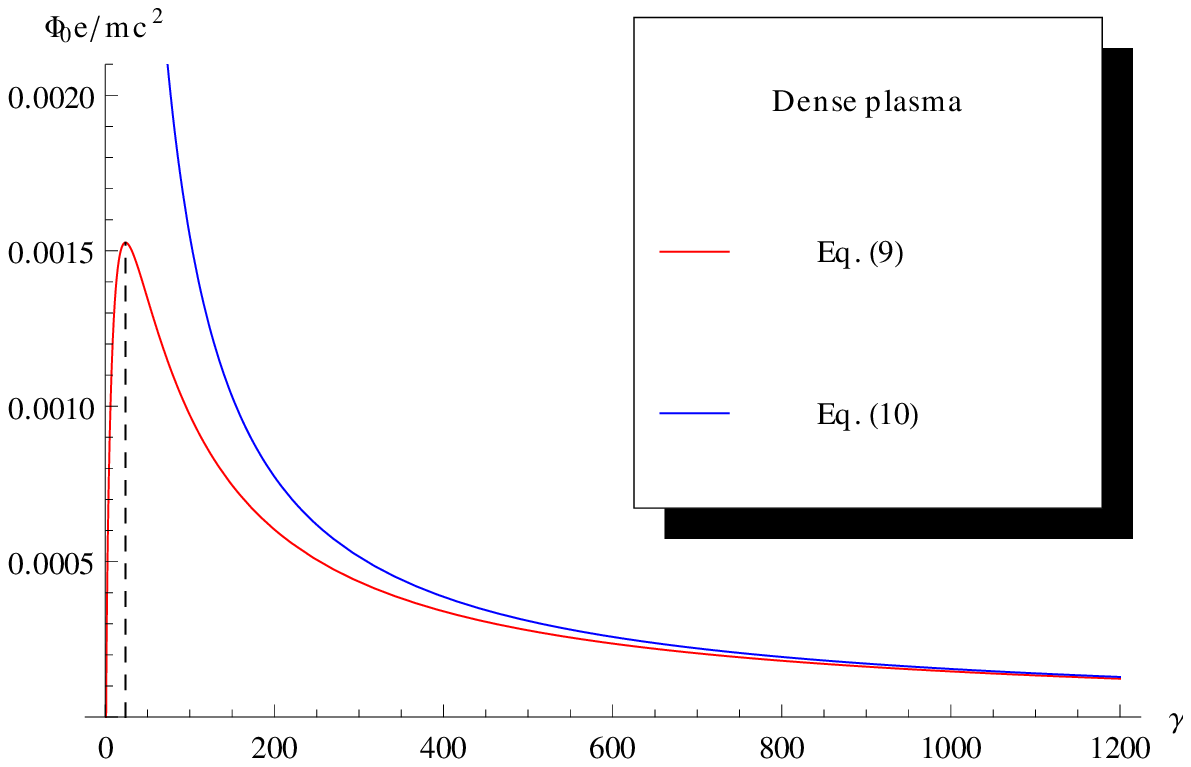}~b)}\\
\subfigure
{\includegraphics[height=6.8 cm,width=7.7cm]{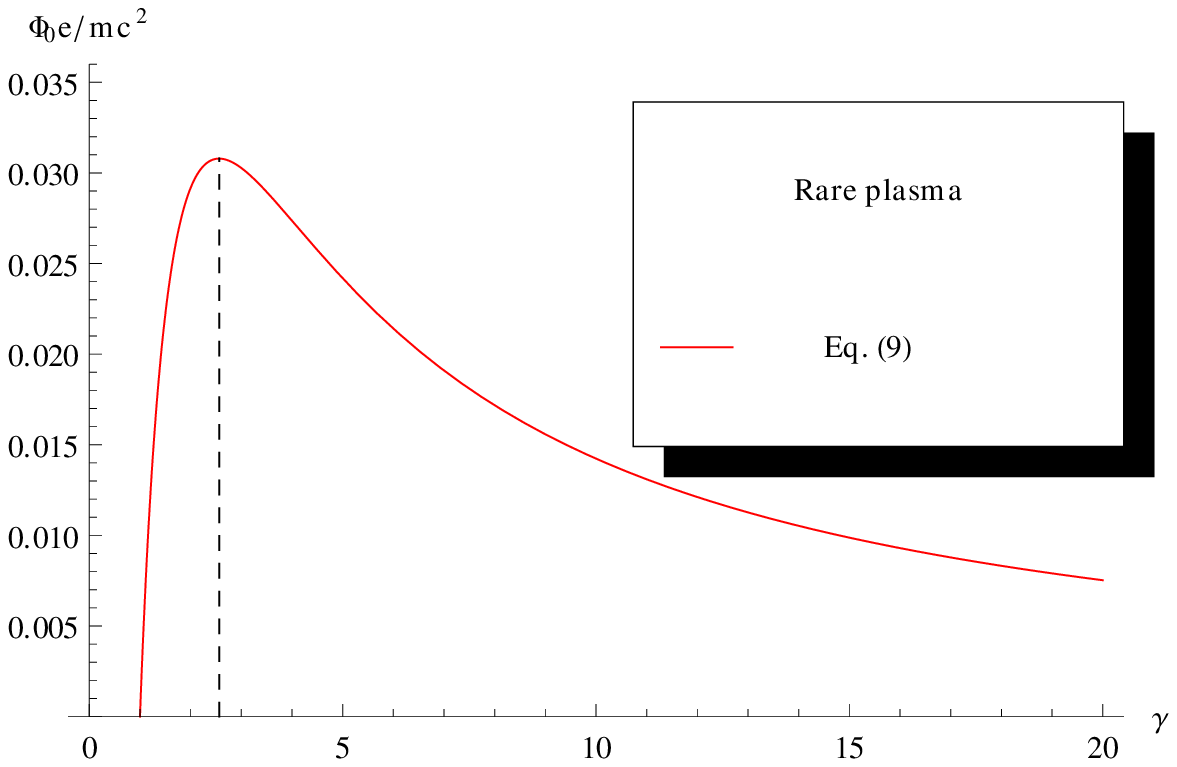}~c)}
\subfigure
{\includegraphics[height=6.8 cm,width=7.6cm]{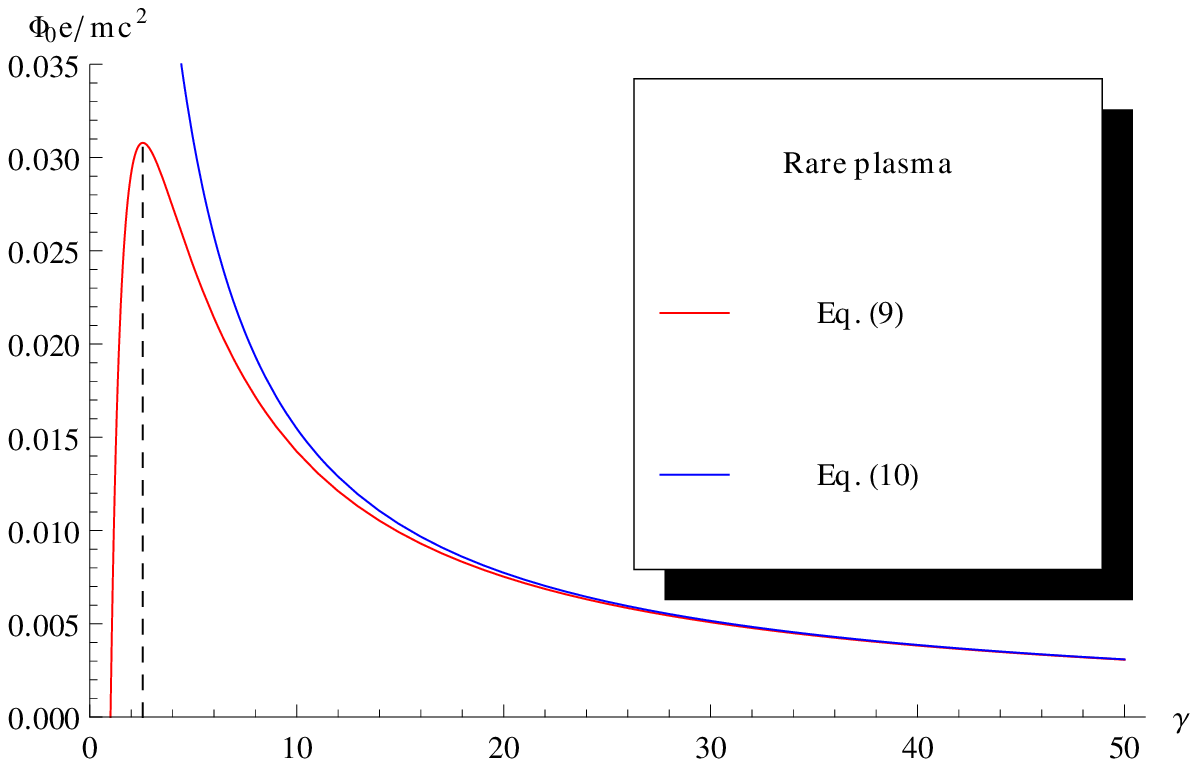}~d)}
\caption{a) The relative stationary saturation amplitude of the plasma wake potential generated by the electron bunch Eq. (\ref{8}) for a), b) dense (\ref{5a}) at $n_{b}=2\cdot 10^{12} cm^{-3}$,  $n_p=10^{16} cm^{-3}$, $\gamma_0=24$ and c), d) rare (\ref{5b}) plasmas at $n_{b}=2\cdot 10^{12} cm^{-3}$,  $n_p=6 \cdot 10^{12} cm^{-3}$, $\gamma_0=2.6$. In b) and d) the relative stationary saturation amplitude in  the ultrarelativistic limit (\ref{9}) is presented for comparison.}
\label{Fig:910}
\end{figure}

Let us start our analysis with the ultrarelativistic electron bunch of density $ n_b$     and  
velocity $\vec{u}$, noting that

\begin{equation}
\label{1}
 \gamma = \frac{1}{\sqrt {(1-u^2/c^2)} } >> 1.
\end{equation} 
\indent Such a bunch interacts with the plasma of density $n_p>>n_b$ and generates the plane wave $E=E_0 \exp(-i\omega t + i \vec k \cdot \vec r)$. We choose an axis Z directed along the velocity of the bunch $\vec u$,  and put the external field  as absent \footnote{ In ultrarelativistic bunches bunch divergence can be neglected because when the electron bunch is ejected into the dense plasma, within the time $t\sim 1/\omega_p$ the neutralization of the bucnh charge occurs prohibitting  the bunch divergence \cite{4}.}, disperse relation describing  the amplification of  a plasma wave with the help of a bunch  can be written as following    \cite{3} :
\begin{equation}
\label{2}
(k^2 c^2 -\omega^2+ \omega_p^2+\omega_b^2\gamma^{-1})\left(1-\frac{\omega_p^2}{\omega^2} -\frac{\omega_b^2\gamma^{-3}}{(\omega-k_z u)^2} \right)-\frac{k_{\perp}^2 u^2}{\omega^2} \frac{\omega_p^2\omega_b^2\gamma^{-1}}{(\omega-k_z u)^2}=0,
\end{equation} 
here $\omega_p= \sqrt{4\pi e^2 n_p/m}$ - Langmuir plasma electron frequency, $k_z$  and $k_\perp$ are the logitudinal (directed along the velocity of the bunch $\vec u$ ) and transverse components  of the wave vector  $\vec k$ .\\
	\indent The general solution of the equation  (\ref{2}) $\omega(\vec k)$ with the positive imaginary part $(\Im m \omega >0)$
\begin{equation}\label{3}
\begin{gathered}
\omega=k_z u(1+\delta)=\omega_p(1+\delta) \hfill \\
\delta=\frac{-1+i \sqrt 3}{2} (\frac{n_b}{2n_p} \frac{1}{\gamma})^{1/3}(1-\frac{u^2}{c^2} \frac{\omega_p^2}{\omega_p^2 +k_\perp^2 u^2} )^{1/3}.
\end{gathered}
\end{equation}
Since the value $\delta$ (increment), determining the temporary growth of the plasma wave amplitude generated by the bunch, depends on $k_\perp$, then in this way it depends on the transverse  size $k_\perp$ of the considered plasma-bunch system. For a start we would like to be far away from the concrete magnitudes of the concrete system parameters and consider two opposite limitting cases: 
\begin{itemize}
\item[A]. In a case of dense plasma when 
 \begin{equation}
\label{4}
\omega_p^2>>k_\perp^2 u^2
\end{equation}
Then,  the value $\delta$  is given by the expression
\begin{equation}
\label{5a}
\delta=\frac{-1+i \sqrt 3}{2} \left(\frac{n_b}{2n_p}\right)^{1/3} \frac{1}{\gamma}.
\end{equation}
\item[B]. In a case of rare plasma when the inverse to (\ref{4}) inequality is satisfied then the value $\delta$ becomes
\begin{equation}
\label{5b}
\delta=\frac{-1+i \sqrt 3}{2} \left(\frac{n_b}{2n_p} \frac{1}{\gamma}\right)^{1/3}.
\end{equation}

\end{itemize}
Let us outline again that while deriving the equations (\ref{3}-\ref{5b}) we took into account that value $\delta$ is negligibly small what is sustained by the inequalities  $n_b<< n_p$ and  $\gamma^2 >> 1$.\\
\indent We would like to stress that for the development of instability and plasma wave growth it is necessary that the following constraint for a system length must be satisfied:
  \begin{equation}
   \label{6}           
       L> \frac{u}{\Im m\delta \omega_p}       
 \end{equation}  
 being at the same time limited due to the bunch divergence to the following constraint :
                
    \[  L < \frac{u\sqrt{\gamma}}{\omega_b} = \frac{u\sqrt{\gamma}}{(n_b/n_p)^{1/2}\omega_p}.\]
\indent  Let us now estimate the maximum amplitude of the generated by the bunch wake, so called the amplitude of saturated instability. For this, we choose the wake as a frame of reference moving with the speed $\omega/k$ with respect to the laboratory frame of reference  (plasma in a rest). In accordance with the Lorentz transformations the speed of bunch electrons in the chosen frame will be 
 \begin{equation}
   \label{7} 
 u_1=-\frac{u \Im m\delta \gamma^2}{1- \frac{2u^2}{c^2} \Im m\delta \gamma^2}=- \frac{u \Im m\delta \gamma^2}{1- 2\Im m\delta (\gamma^2-1)}.
 \end{equation}  
 From Eq. (\ref{7}) follows that in the relativistic limit when $ (\delta\gamma^2 << 1)$ this speed is small $u_1 \approx -u \delta \gamma^2<< u$, whereas in the ultrarelativistic limit $ (\delta\gamma^2 >> 1)$ the speed is high $ u_1 \approx u/2$.\\
	\indent It is obvious that the saturation of instability can occur when the kinetic energy of electrons, in the wake frame of reference, will become less than the amplitude of the potential  of the plasma wake measured in the same frame. In this case the bunch electrons get trapped by the wake, i.e. there will be no relative motion between the bunch electrons and the wake, thus, no energy exchange between the bunch and wake occurs, the bunch and the wake become stationary.  Hence, the stationary saturation amplitude of the plasma wake potential can be obtained as following
	\begin{equation}
   \label{8}
	\frac{e \Phi_0}{mc^2} =\frac{1}{\gamma}\left\{\frac{1}{\sqrt{1-\frac{u^2 \delta^2 \gamma^4}{c^2 (1- 2\delta (\gamma^2-1))^2}}} -  1  \right\} ,
	\end{equation}  
	here $\Phi_0\gamma$  is measured in the wake frame of reference. In the relativistic limit when $ (\delta\gamma^2 << 1)$ we will get the result already obtained in  \cite{9} for dense plasma Eq. (\ref{5a}), whereas in the ultrarelativistic limit when $(\delta \gamma^2 >> 1)$ we obtain 
	\begin{equation}
   \label{9}
	\frac{e \Phi_0 }{mc^2} \approx \frac{0.154}{\gamma},
	\end{equation}  
	here $\Phi_0$ is measured in the laboratory frame of reference. Eq. (\ref{9}) leads to the following conclusion: the amplitude of the potential of the plasma wake generated by the electron bunch is dependent on the relativity factor ${\Phi_0 }= \frac{0.154 mc^2}{e\gamma}$.  This is demonstrated in the Figure \ref{Fig:910}. We can obeserve a maximum ${{\Phi_0}_{max} } \simeq 770 $  V for dense plasma at $n_p= 10^{16} cm^{-3}$, $\gamma_0\simeq 24$ and ${{\Phi_0}_{max} } \simeq 15.5 \cdot 10^3$ V $\simeq 15.5$ k V rare plasma at $n_p= 6 \cdot 10^{12} cm^{-3}$, $\gamma_{0}\simeq 2.56$, see  \ref{222}. It is quite easy to estimate the amplitude of the electrial field of the wake generated by the bunch. For this in Eq.  (\ref{8}) one needs to take into account that $E_0 \sim k  \Phi_0$  leading to the following equation for the relative energy density of  the electrostatic field 
		\begin{equation}
   \label{10}
	\frac{E_0^2}{8 \pi n_p mc^2 \gamma} \simeq \frac{c^2}{u^2} \frac{1}{2\gamma^3} \left \{\frac{1}{{\sqrt{1-\frac{{u^2 \delta^2 \gamma^4}}{c^2 (1- 2\delta (\gamma^2-1))^2}}}} -1 \right\}^2.
	\end{equation}  
	In the relativistic limit when $(\delta \gamma^2 << 1)$ this equation turns into the result obtained in \cite{9} for dense plasma Eq.  (\ref{5a}), whereas in the ultrarelativistic limit when $(\delta \gamma^2 >> 1)$ we get 
	\begin{equation}
   \label{11}
	\frac{E_0^2}{8 \pi n_p mc^2 \gamma} \simeq  \frac{0.012}{\gamma^3} 
	\end{equation}  
	We can readily see that with an increase of relativity the amplitude of the electric field of the wake generated by the electron bunch in its turn is dependent on the relativity $\gamma$  as it is shown in the Figure \ref{Fig:910} and is also defined in the ultrarelativistic limit by the plasma  density $n_p$, $E_0 \simeq 0.024\sqrt {4 \pi n_p mc^2 }/\gamma$. For example, for dense plasma  at $\gamma_0\simeq 24$, $n_p=10^{16} cm^{-3}$, we have ${E_0}_{max} \simeq 14.5 \cdot 10^6 $ V/cm $\simeq 14.5 $ M V/m, whereas for rare plasma at $n_p= 6 \cdot 10^{12} cm^{-3}$, $\gamma_{0}\simeq 11.6$ we have ${E_0}_{max} \simeq 7.2 \cdot 10^6 $ V/cm $\simeq 7.2 $ M V/m.
	
\section{ Ultrarelativistic proton bunch}

As it has been mentioned above, in works \cite{7, 8} the possibility of generation of high power wakefields  (proton-bunch-driven plasma-wakefield acceleration) of terra-watt amplitude using the  ultrarelativistic  proton bunches of giga-electron-volt energy was introduced.   The idea is that to slow the protons down using the plasma wakefield is much harder than the electrons and as a result the significantly higher wakefield amplitude can be gained in comparison with that produced by the electron bunch. \\
	\indent The disperse relation (\ref{2}) can be easily generalized for the case of plasma-proton-bunch interaction. Correspondingly, we need to make the substitution $n_b  \to n_{bi}$ ($n_{b}$ - electron bunch density), i.e. substitute the $n_{b}$ by the proton bunch density $n_{bi}$ multiplied by the ratio of electron mass $m$  to the ion mass $M$, $m/M$. \\
\indent  Having solved Eq. (\ref{2}), we write the solution generalizing Eq. (\ref{3}) for the proton bunch down
	 \begin{equation}\label{12}
\begin{gathered}
\omega=k_z u(1+\delta_1)=\omega_p(1+\delta_1) \hfill \\
\delta_1=\frac{-1+i \sqrt 3}{2} (\frac{n_b}{2n_p}\frac{m}{M}\frac{1}{\gamma})^{1/3}(1-\frac{u^2}{c^2} \frac{\omega_p^2}{\omega_p^2 +k_\perp^2 u^2} )^{1/3},
\end{gathered}
\end{equation}
	  here $\delta_1$ differs from $\delta$, from  Eq. (3), in $n_b \to n_{bi}\frac{m}{M}$.
	
	\begin{itemize}
\item[A]. In the dense plasma when the constraint (\ref{4}) is satisfied  (cmpr. with (\ref{5a})) the value $\delta_1$   takes the following view:
\begin{equation}
\label{13a}
\delta_1=\frac{-1+i \sqrt 3}{2} \left(\frac{n_{bi}}{2n_p}\frac{m}{M}\right)^{1/3} \frac{1}{\gamma}.
\end{equation}
\item[B]. In the rare plasma when the inverse to (\ref{4}) inequality is satisfied then the value $\delta_1$ (cmpr. with (\ref{5b}))  becomes
\begin{equation}
\label{13b}
\delta_1=\frac{-1+i \sqrt 3}{2} \left(\frac{n_{bi}}{2n_p}\frac{m}{M} \frac{1}{\gamma}\right)^{1/3}.
\end{equation}
\end{itemize}
	\indent  We would like to point out again that as in the case for the electron bunch the development of proton bunch instability and plasma wake growth can occur when the following constraint similar to Eq. (\ref{6}) for a system length must be satisfied
	 \begin{equation}
\label{14}
	 L> \frac{u}{\Im m\delta_1 \omega_p}.
\end{equation}
However, the system length must be limited as well in order to neglect the proton bunch divergence due to the proton charges repulsion 
\[L < \frac{u\sqrt{\gamma}}{\omega_{bi}} = \frac{u\sqrt{\gamma}}{(\frac{n_{bi}}{n_p} \frac{m}{M})^{1/2}\omega_p}.\]

\begin{figure}
\subfigure
{\includegraphics[height=6.8 cm,width=7.7cm]{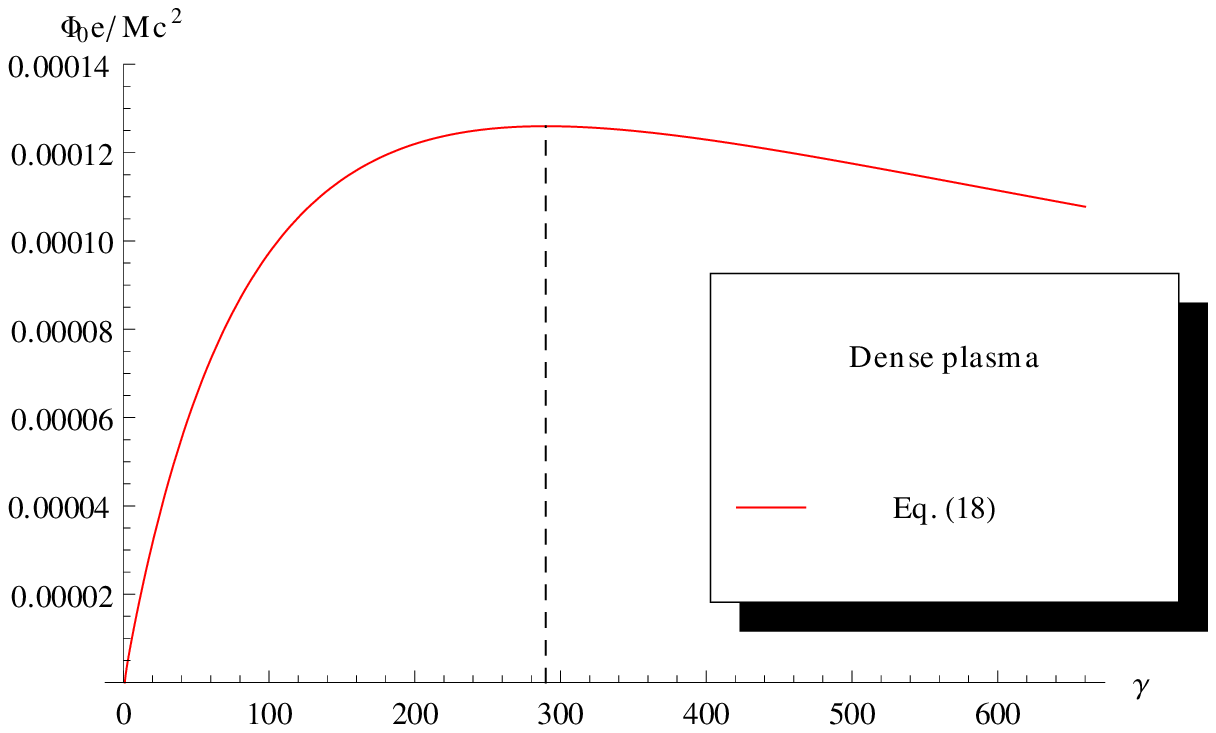}~a)}
\subfigure
{\includegraphics[height=6.8 cm,width=7.6cm]{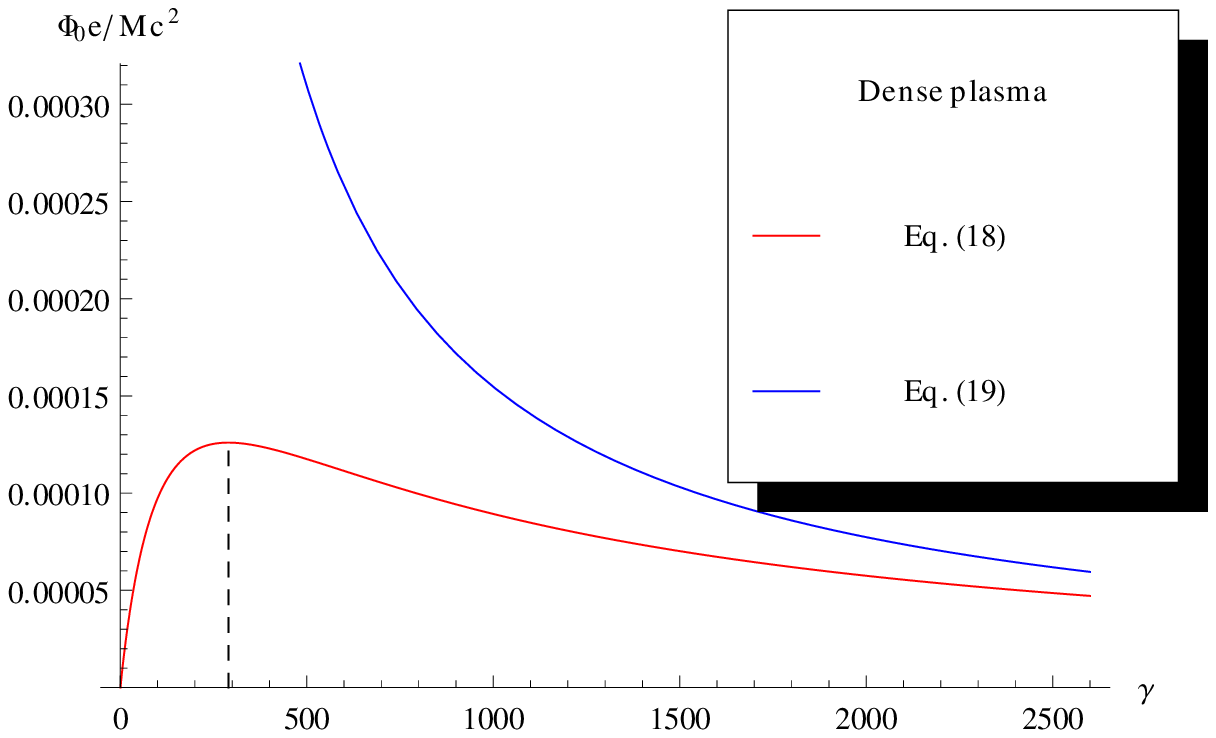}~b)}\\
\subfigure
{\includegraphics[height=6.8 cm,width=7.7cm]{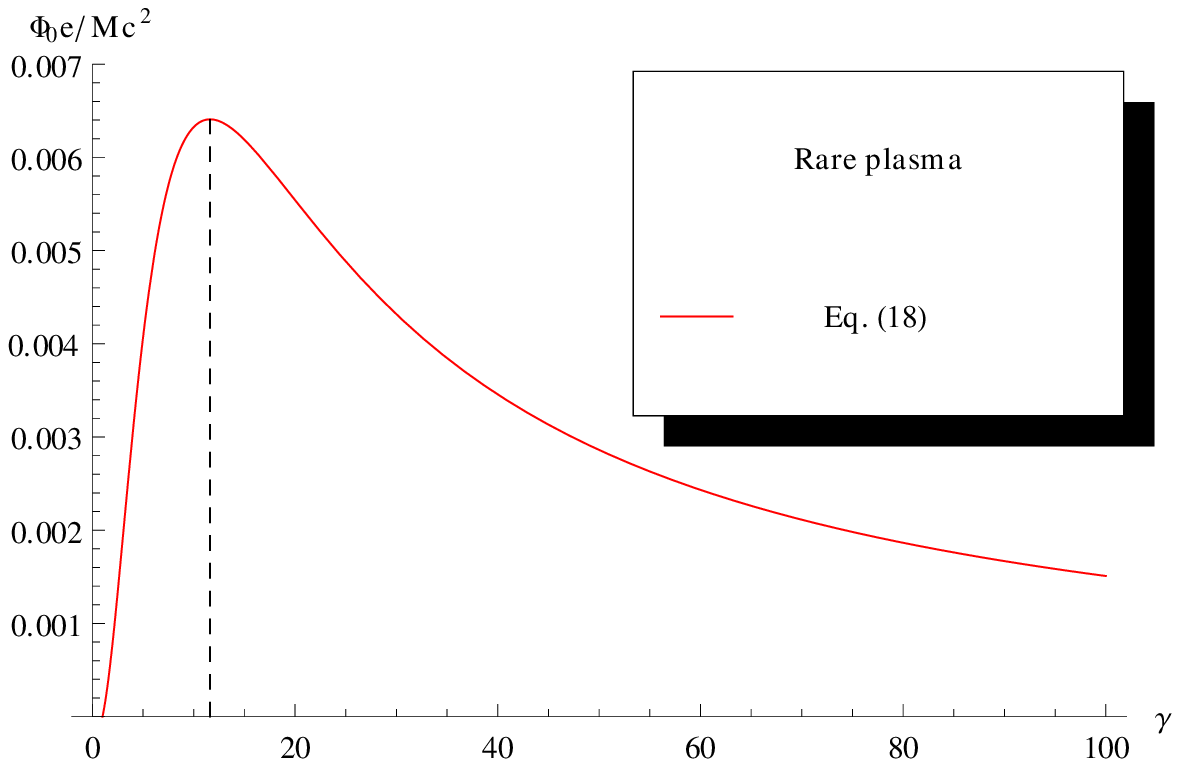}~c)}
\subfigure
{\includegraphics[height=6.8 cm,width=7.6cm]{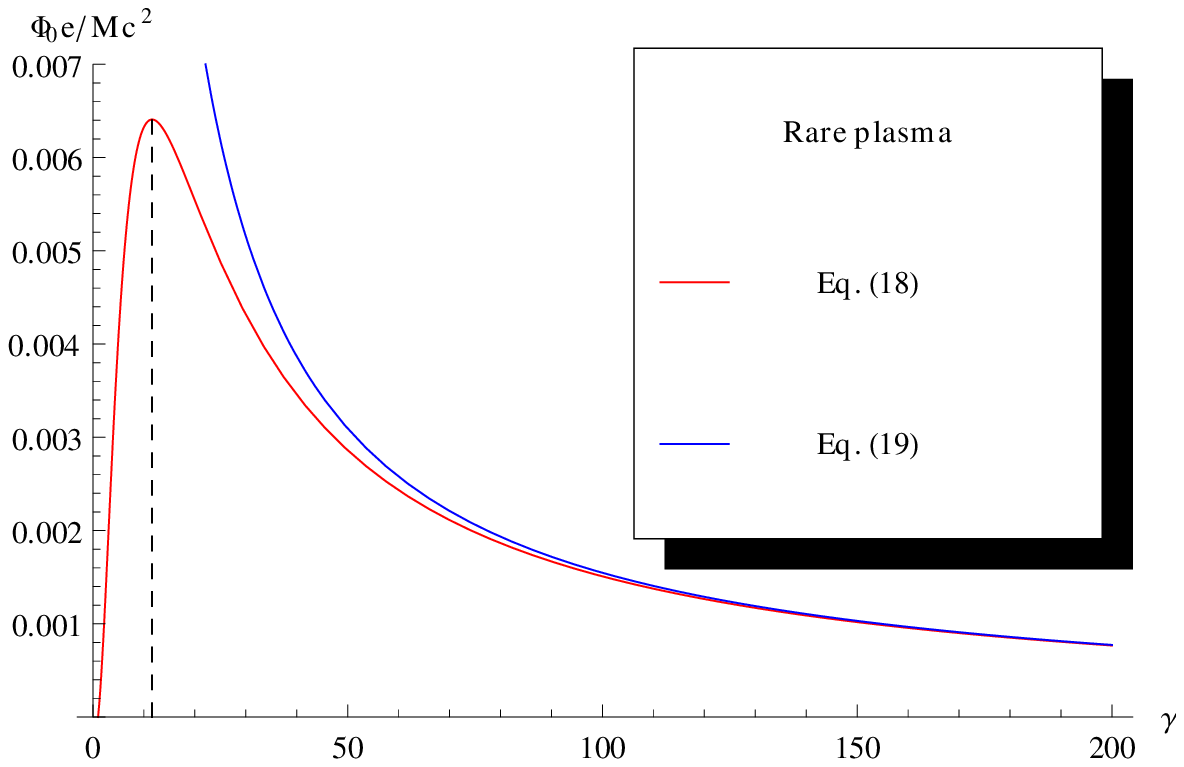}~d)}
\caption{a) The relative stationary saturation amplitude of the plasma wake potential generated by the proton bunch Eq. (\ref{16}) for a), b) dense (\ref{13a}) at $n_{b}=2\cdot 10^{12} cm^{-3}$,  $n_p=10^{16} cm^{-3}$, $\gamma_0=289$ and c), d) rare (\ref{13b}) plasmas at $n_{b}=2\cdot 10^{12} cm^{-3}$,  $n_p=6 \cdot 10^{12} cm^{-3}$, $\gamma_0=11.6$. In b) and d) the relative stationary saturation amplitude in  the ultrarelativistic limit (\ref{17}) is presented for comparison.}
\label{Fig:920}
\end{figure}

Similarly, we estimate the maximum amplitude of the generated by the proton bunch wake, the amplitude of saturated instability. For this, we choose again the wake as a frame of reference moving with the speed $\omega/k$ with respect to the laboratory frame of  reference. In this case,  the speed of bunch protons in this frame (cmpr. with (\ref{7})) will be 	  
\begin{equation}
\label{15}
u_1=-\frac{u \Im m\delta_1 \gamma^2}{1- \frac{2u^2}{c^2} \Im m\delta_1 \gamma^2}=- \frac{u \Im m\delta_1 \gamma^2}{1- 2\Im m\delta_1 (\gamma^2-1)}. 
\end{equation}
From Eq. (\ref{15}) we get the following result:  in the relativistic limit when $ (\delta\gamma^2 << 1)$  this speed is small $u_1 \simeq -u \delta_1\gamma^2 << u$, whereas in the ultrarelativistic limit ($ \delta\gamma^2 >> 1$) the speed is high  $u_1 \simeq u/2 $. It is worth to point out that ultrarelativity of the proton bunch can be reached at much higher proton energies : $\gamma^2$  should be at least one order of magnitude higher compared to that of the electron bunch. For example, when for electron bunch - $n_{b}/n_p \sim 10^{-3}$ , the  ultrarelativity can be reached at $ \gamma \geq 10$ (electron bunch energy is higher than 5 MeV), whereas for the proton bunch at the same density ratio - only at $ \gamma \geq 50$ (proton bunch energy should be higher than 44 GeV).\\
\indent	By analogy with Eq. (\ref{8}), the stationary saturation amplitude of the plasma wake potential, produced by the proton bunch  can be obtained as following
\begin{equation}
   \label{16}
	\frac{e \Phi_0}{Mc^2} =\frac{1}{\gamma}\left\{\frac{1}{\sqrt{1-\frac{u^2 \delta_1^2 \gamma^4}{c^2 (1-2 \delta_1 (\gamma^2-1))^2}}} -  1  \right\}, 
	\end{equation}  
here $\Phi_0\gamma$  is measured in the wake frame of reference. Correspondingly, in the ultrarelativistic limit when $\delta_1 \gamma^2 >> 1$ we get an estimate  similar to the Eq. (\ref{9})  
\begin{equation}
   \label{17}
	\frac{e \Phi_0 }{Mc^2} \approx \frac{0.154}{\gamma},
	\end{equation}  
	here $\Phi_0$ is measured in the laboratory frame of  reference. Eq. (\ref{17}) leads the following conclusion: the saturation amplitude of the potential of the plasma wake generated by the ultrarelativistic proton bunch is also dependent on the relativity factor  ${\Phi_0 }= \frac{0.154 Mc^2}{e\gamma}$. This can be observed in the Figure \ref{Fig:920}. As we can see it posesses a maximum. We have estimated it for dense plasma at $n_p= 10^{16} cm^{-3}$, $\gamma_{0}\simeq 289$ leading to ${{\Phi_0}_{max} } \simeq 1.1 \cdot 10^5$ V $\simeq 0.1$ M V and rare plasma at $n_p= 6 \cdot 10^{12} cm^{-3}$, $\gamma_{0}\simeq 11.6$ leading to ${{\Phi_0}_{max} } \simeq 5.74 \cdot 10^6$ V $\simeq 5.74$ M V, see \ref{222}. \\
\indent Again, it is quite easy  to estimate the amplitude of the electrial field of the wake generated by the proton bunch. For this in Eq. (\ref{16}), taking into account that $E_0\sim k\Phi_0$,  we obtain the following equation for the relative energy density of the electrostatic field 
\begin{equation}
   \label{18}
	\frac{E_0^2}{8 \pi n_p M c^2 \gamma} \simeq \frac{c^2}{u^2} \frac{1}{2\gamma^3} \left \{\frac{1}{{\sqrt{1-\frac{{u^2 \delta_1^2 \gamma^4}}{c^2 (1- 2\delta_1 (\gamma^2-1))^2}}}} -1 \right\}^2
	\end{equation} 
	leading in the ultrarelativistic limit for the proton bunch when $\delta_1 \gamma^2 >> 1$ to (cmpr. with (\ref{11})) 
	\begin{equation}
   \label{19}
	\frac{E_0^2}{8 \pi n_p M c^2 \gamma} \simeq  \frac{0.012}{\gamma^3} . 
	\end{equation} 
\indent 	We can readily see that the amplitude of the potential as well as that of the electric field of the wake generated by the proton bunch in its turn is defined by the relativity factor as it is shown in the Figure \ref{Fig:920} and plasma density $n_p$, namely $E_0  = 0.024\sqrt {4 \pi n_{p} M c^2 }/\gamma$.  For ex., for dense plasma at $n_p= 10^{16} cm^{-3}$, $\gamma_{0}\simeq 289$  we have ${E_0}_{max}  \simeq 5 \cdot 10^5$ V/cm $\simeq 50$ M V/m, for rare plasma at $n_p= 6 \cdot 10^{12} cm^{-3}$, $\gamma_{0}\simeq 11.6$  we have ${E_0}_{max}  \simeq 6.3 \cdot 10^5$ V/cm $\simeq 63$ M V/m .
	
\section{Results and Discussions}
On the basis of the conducted analysis of the resonance Cherenkov interaction of the electron as well as proton bunches with plasma, we can make the following conclusion: the wake amplitude growth produced by the bunches gets saturated with an increase of bunch energy at   a quite high level. The highest amplitude of the electric wakefield is produced by the electron bunch in dense plasma at $n_p= 10^{16} cm^{-3}$  and is of order $14.5$ M V/m, whereas that produced by the proton bunch is the highest in the rare plasma at $n_p= 6 \cdot 10^{12}  cm^{-3}$ and is of order  $63$ M V/m.  These magnitudes are less than those gained with the help of contemporary quite powerful  pulse lasers ($10^{15}$ W/cm$^{2}$).\\
\indent In conclusion it must be noticed that nonlinear plasma Lengmuir wave is unstable and, as it was shown in \cite{Akh}, it breaks down when
           \[e \Phi_0 > \{M,m\}c^2\gamma_w,\]
where $\gamma_w=1/\sqrt{1-u_{ph}^2/c^2}$ and $u_{ph}$ is the wave phase velocity. In our case $u_{ph}=u$ and $\gamma_w=\gamma$. From the results of our calculations follows that considered resonance Cherenkov instabilities arise as for electron as well as for proton beams when the plasma waves amplitudes are much less than the breakdown threshholds.
\appendix

 \section{Estimation of the maximum value of the wake field potential}
\label{222}
In order to determine the maximum value of the $\Phi_0$ let us represent the Eq. (\ref{8}) in the following independent on the type of interaction either electron-plasma or proton-plasma:
\begin{equation}
\label{888}
\frac{e \Phi_0}{mc^2} =(\frac{\alpha}{Z})^{1/n}\left\{\frac{1+2Z}{\sqrt{1+4Z+3Z^2}}-1\right\} ,
\end{equation}  
where $Z=\delta_{\{\:\},1}\gamma^2=-\alpha\gamma^n$, $n=1$ corresponds to the dense plasma and $n=5/3$ - the rare plasma,  $\alpha=\frac{-1}{2}\left(\frac{n_{bi}}{2n_p}\frac{m}{M}\right)^{1/3}$ for proton bunch and $\alpha=\frac{-1}{2} \left(\frac{n_{b}}{2n_p}\right)^{1/3}$ for electron bunch.\\
\indent This Eq. (\ref{888}) has a peak at $Z=0.55546$ for dense plasma (n=1) and the parameters $\alpha$ and $\gamma$ should be changed according to the realtion
\begin{equation}
   \label{889}
0.55546=\alpha\gamma.
\end{equation}  
  For rare plasma (n=5/3) at $Z=1.35695$ we have
 \begin{equation}
   \label{900}
1.35695=\alpha\gamma^{5/3}.
\end{equation}  
These values correspond to the maxima for dense plasma 
\begin{equation}
   \label{999}
\frac{e \Phi_0}{\{M,m\}c^2}=-0.06576 \alpha 
\end{equation}  
and  for rare plasma 
 \begin{equation}
   \label{1000}
\frac{e \Phi_0}{\{M,m\}c^2}=-0.0618 \alpha^{3/5}.
\end{equation}  
 From Eqs. (\ref{999}) and (\ref{1000}) follows that we should increase $\alpha$ in such a way that the constraints (\ref{4}), (\ref{5a}), (\ref{5b}) and (\ref{13a}), (\ref{13b}), $\delta_{\{\:\},1}<< 1$, $\gamma^2>>1$ are not violated. As one can readly see the highest maximum saturation amplitude of the plasma wake potential corresponds to the rare plasma, whereas the highest maximum electric wakefield is in addition defined by the plasma density. 
  


\bibliographystyle{model1a-num-names}







\end{document}